\documentclass[12pt]{article}
\usepackage{amsmath}
\usepackage{graphicx,psfrag,epsf}
\usepackage{enumerate}
\usepackage{natbib}
\usepackage{url} % not crucial - just used below for the URL 
\usepackage{amsthm} %package that handles proofs etc
\usepackage{float} %handles figures as float objects
\usepackage{import} % handles imports
\usepackage{booktabs}
\usepackage[toc, page] {appendix}

\title{Optimal Allocation of Limited Funds in Quadratic Funding}

\author{ Ricardo A. Pasquini \thanks{Facultad de Ciencias Empresariales, Universidad Austral, Mariano Acosta 1611, B1630FHB Pilar, Buenos Aires. The
views expressed in the paper do not necessarily reflect those of the institutions I am affiliated with.. \href{mailto:rpasquini@austral.edu.ar}{rpasquini@austral.edu.ar}} }

\providecommand{\keywords}[1]{\textbf{\textit{Keywords---}} #1}
\providecommand{\jelcodes}[1]{\textbf{\textit{JEL codes---}} #1}

\providecommand{\tightlist}{%
  \setlength{\itemsep}{0pt}\setlength{\parskip}{0pt}}

\newtheorem{theorem}{Theorem}

\newtheorem{observation}[theorem]{Observation}

\usepackage[breaklinks=true]{hyperref}
\hypersetup{colorlinks=true,%
citecolor=blue,%
filecolor=blue,%
linkcolor=blue,%
urlcolor=blue}
\usepackage{url}

\begin{document}

\def\spacingset#1{\renewcommand{\baselinestretch}%
{#1}\small\normalsize} \spacingset{1}

\maketitle

\begin{abstract}
We examine the allocation of a limited pool of matching funds to public
good projects using Quadratic Funding. In particular, we consider a
variation of the Capital Constrained Quadratic Funding (CQF) mechanism proposed by
Buterin, Hitzig and Weyl (2019) where only funds in the matching pool are distributed among projects. We show that this mechanism achieves a
socially optimal allocation of limited funds.
\end{abstract}

    \keywords{quadratic funding, public goods}

    \jelcodes{D47, D61, D71, H41}

%\newpage

\spacingset{1.45} % DON'T change the spacing!

\hfill\break

Quadratic Funding (QF) is a target matching funds allocation rule for
public goods financing proposed by Buterin, Hitzig, and Weyl (2019)
(BHW). Denoting by \(c_i^p\) the amount committed by an individual
contributor to project \(p\), the QF rule proposes funding project \(p\)
in the amount \(F^p\) according to the formula that squares the sum of
the square roots of contributions from all contributors (i.e.,
\(F^p=(\sum_{i\in I}\sqrt{c_i})^2\)). BHW shows that the QF rule has a
powerful theoretically property. Under this mechanism individual
contributors choose (in a decentralized way) a socially efficient
allocation of funds. Indeed, a social planner would balance the marginal
cost of an additional dollar invested in a project against the sum of
the marginal benefits for all the community members, and this is what QF
achieves in its decentralized design\footnote{See Buterin, Hitzig, and
  Weyl (2019) (BHW), Proposition 3, pp.~5175}.

In practice, implementing the QF rule fully will probably prove
difficult. This is because matching funds requirements scale rapidly
(quadratically), and available matching funds (typically provided by
external donors) are limited. In \emph{Gitcoin Grants} (a website that
has implemented QF to fund projects), for example, requirements reach
the total matching pool within the first days of a round (Pasquini
2020). As a consequence, what is actually implemented in practice is a
variant of QF, named by BHW the \emph{Capital Constrained QF} (CQF). By
this rule what a project obtains is not the full amount from the QF
rule, but a linear combination between the target QF and actual funds
committed by individual contributors.

Once we recognize that matching funds will be limited in general, it is
worth consider if CQF is the best rule for the allocation of a
\emph{limited pool of funds}. An optimal allocation of limited funds
should follow the condition that next dollar invested by the mechanism
should go to the project with the highest marginal social benefit.
Therefore, investments should equalize marginal social benefits across
projects. It turns out that this is not perfectly achieved by CQF.
Differences between marginal social benefits across projects will be
larger: i) the lesser funds are available in the matching pool relative
to matching requirements, ii) the higher the variability in the
supporting preferences across projects (e.g., more equally invested
projects imply higher marginal benefits than more concentrated
projects), and iii) the higher the number of contributors (Pasquini
2020).

Here we propose examining the allocation of limited funds to public good
projects using a slight variation of the CQF rule. In this version:

\begin{itemize}
\tightlist
\item
  Projects do not receive the funds committed directly by contributors.
  They do receive a percentage of the total funds a project should
  receive according to the QF rule. In other words, only the funds in
  the pool of matching funds are distributed.
\item
  Individual contributions are kept by the mechanism, and could be
  summed to the matching pool of funds in subsequent financing rounds.
\end{itemize}

We show that, under perfect information, this mechanism tends to a
social optimal allocation of \emph{limited funds}, namely it equalizes
marginal social benefits across projects. Individual contributors have
incentives to contribute (and change the allocation of funds), if the
marginal social benefit from a project of interest is greater than the
(public goods) weighted marginal social benefit they would receive from
all (other) projects. This incentive tends to disappear as marginal
social benefits tend to equalize across projects.

Notice that because of quadratic funding, individual contributors fully
internalize the decisions on the contributions from the rest of the
contributors (i.e., the community). This is the main innovation from QF.
In this particular case, differently from the BHW's CQF, by removing the
effect of private commitments to projects, marginal social benefits tend
to equalize without biases.

With limited funds, we can think of contributors as paying for the right
to distribute the matching pool of funds in favor of their project of
interest, but without the security of fully delivering their committed
funds to their project of interest. With enough matching funds, the QF
rule determines an amount to a project that is always greater than
individual commitments, but with limited funds, in this design, the
mechanism could deliver a share of those.

We will also note that the price contributors need to pay for
distributing matching funds towards their projects of interest increases
as the financing round progresses. This is because this cost is higher
the higher are QF-rule fund requirements relative to actual funds in the
matching pool. So any contribution to the mechanism increases this
price. As a result, contributors eventually cease to invest in
reallocating funds.

\hypertarget{mechanism}{%
\section*{The mechanism}\label{mechanism}}
\addcontentsline{toc}{section}{The mechanism}

Assume that \(p\in P\) indexes public good projects competing to receive
funding. Also \(i \in I\) indexes individual contributors. An individual
\(i\) supports a project \(p\) by committing an amount of money
\(c_i^p\). In addition, assume there is a pool of funds provided by
donors, that we will denote \(D\), and that will be used to match
individual contributions.

The mechanism promises, for each project \(p\), an investment of: \[
F^p=\bigg\{\begin{matrix}
  F^{p,QF}, &  \text{if } \sum_p F^{p,QF}\leq D\\
  \frac{F^{p,QF}}{\sum_p F^{p,QF}}D, & \text{if }\sum_p F^{p,QF}> D
  \end{matrix}
\] Where \(F^{p,QF}\) denotes the target amount of funds according to
the QF Rule (i.e.,\(F^{p,QF}=(\sum\sqrt{c_i^p})^2\) ).

If contributions to projects are such that \(\sum_p F^{p,QF}\leq D\),
then the mechanism is essentially the \emph{capital unconstrained}
version of QF that, as already discussed, leads to a socially efficient
outcome (See BHW).

In case \(\sum_p F^{p,\text{QF}}\geq D\), the mechanism is restricted to
distribute \(D\) across projects, according to the shares
\(\frac{F^p}{\sum_pF^p}\)\(p\in P\) . This essentially implies that
projects compete for a fixed amount of funding \(D\). By investing in a
project \(p\), contributors change the share of funds \(p\) will
receive, at the cost of invested funds (n.b. increasing invested funds
do not change \(D\)).

Notice that contributions \(c_i^p\) are not distributed by the
mechanism. Since in practice QF rounds are regularly repeated, funds
raised by the mechanism in a given round could be used as part of a the
pool of matching funds in a subsequent round.

\hypertarget{contributorproblem}{%
\subsection*{Individual contributor problem}\label{contributorproblem}}
\addcontentsline{toc}{subsection}{Individual contributor problem}

As in Buterin, Hitzig, and Weyl (2019), let \(V_i^p\) be the
currency-equivalent utility a citizen \(i\) receives from a public good
\(p\). Utilities from different public goods are assumed to be
independent, and it is assumed a setting of complete information.

A contributor \(i\) chooses how much to contribute to each project
\(p\in P\). We denote such decisions as \(\{c_i^p\}_{p\in P}\). We will
assume that contributors assume there will be limited matching funds
available. Contributor's \(i\) optimization is defined by \[
\max_{\{c_i^p\}_{p\in P}} \sum_{p' \in P}V_i^{p'}\bigg(\frac{D}{\sum_{p'\in P}F^{p'\text{,QF}}}F^{p'\text{,QF}}\bigg)-c_i^{p'}
\]

Notice that (differently from BHW) here we consider the sum of utilities
across all projects. Because there are limited funds, contributing
decisions cannot be independently considered.

The F.O.C. for \(c_i^{p}\) is \[
\begin{aligned}
V_i^{'p}(.)\bigg(\frac{\sum_i\sqrt{c_i^p}}{\sqrt{c_i^p}}\frac{D}{\sum_{p'\in P}F^{p'\text{,QF}}}-\frac{D}{(\sum_{p'\in P}F^{p'\text{,QF}})^2}F^{p,\text{QF}}\frac{\sum_i\sqrt{c_i^p}}{\sqrt{c_i^p}}\bigg)-1+\\ 
\sum_{p'\in P,\ p' \neq p}V_i^{'p'}(.)\bigg(-\frac{D}{(\sum_{{p'}\in P}F^{p'\text{,QF}})^2}\frac{\sum_i\sqrt{c_i^p}}{\sqrt{c_i^p}}F^{p'\text{,QF}}\bigg)=0
\end{aligned}
\] The first term gives the direct effect of contributing in \(p\) plus
an indirect effect caused by the fact that increasing \(c^p_i\) also
increases the pool of matching requirements
\(\sum_{p'\in P}F^{p'\text{,QF}}\). The second term gives the indirect
effect of increasing the pool of matching requirements in all the
remaining projects (\(p' \neq p\)).

Indirect effects can be grouped \[
V_i^{'p}(.)\bigg(\frac{\sum_i\sqrt{c_i^p}}{\sqrt{c_i^p}}\frac{D}{\sum_{p'\in P}F^{p'\text{,QF}}}\bigg)+\\ 
\sum_{p'\in P}V_i^{'p'}(.)\bigg(-\frac{D}{(\sum_{{p'}\in P}F^{p'\text{,QF}})^2}\frac{\sum_i\sqrt{c_i^p}}{\sqrt{c_i^p}}F^{p'\text{,QF}}\bigg)=1
\] Then, grouping common factors yield \begin{equation}
\frac{\sum_i\sqrt{c_i^p}}{\sqrt{c_i^p}}\bigg(\frac{D}{\sum_{p'\in P}F^{p'\text{,QF}}}\bigg)\bigg[V_i^{'p}(.)- 
\sum_{p'\in P}V_i^{'p'}(.)\frac{F^{p'\text{,QF}}}{(\sum_{{p'}\in P}F^{p'\text{,QF}})}\bigg]=1
\label{eq:cpo1}\end{equation}

Equation \ref{eq:cpo1} can be interpreted as follows: Contributing to
\(p\) increases the marginal benefit from project \(p\) (i.e.,
\(V_i^{'p}\)), but this is scaled, according to the QF rule, by the
relative contributors of all other individuals (i.e.,
\(\frac{\sum_i\sqrt{c_i^p}}{\sqrt{c_i^p}}\)). Because there are limited
funds, this benefit it is also scaled-down by the ratio of available (to
target) funds (i.e., \(\frac{D}{\sum_{p'\in P}F^{p'\text{,QF}}}\)). In
addition, contributing to project \(p\) now reduces the availability of
funds to all other projects, so there is a negative effect that weights
the marginal disutility of each project by the relative size of target
funding.

Rearranging terms yields \[
V_i^{'p}(.)- 
\sum_{p'\in P}V_i^{'p'}(.)\frac{F^{p'\text{,QF}}}{(\sum_{{p'}\in P}F^{p'\text{,QF}})}=\frac{\sqrt{c_i^p}}{\sum_i\sqrt{c_i^p}}\bigg(\frac{\sum_{p'\in P}F^{p'\text{,QF}}}{D}\bigg)
\] Summing each side of the equation across individuals yields \[
\sum_iV_i^{'p}(.)- 
\sum_i\sum_{p'\in P}V_i^{'p'}(.)\frac{F^{p'\text{,QF}}}{(\sum_{{p'}\in P}F^{p'\text{,QF}})}=\frac{\sum_{p'\in P}F^{p'\text{,QF}}}{D}
\]

Or equivalently \begin{equation}
\sum_iV_i^{'p}(.)-\sum_{p'\in P}\frac{F^{p'\text{,QF}}}{(\sum_{{p'}\in P}F^{p'\text{,QF}})}\sum_iV_i^{'p'}(.)=\frac{\sum_{p'\in P}F^{p'\text{,QF}}}{D}
\label{eq:sumwaverage}\end{equation} It is useful to get an intuition on
Equation \ref{eq:sumwaverage}. Notice that \(\sum_iV_i^{'p}(.)\) is the
sum of the marginal utilities across individuals from public good \(p\).
The second term, is a (public-good sized) weighted average of the sum of
marginal utilities. If the sum of the marginal utilities from \(p\) is
greater than the weighted average from all projects, individuals will be
incentivized to invest in \(p\) (i.e., there will be a reallocation of
limited funds towards \(p\)). But investments in \(p\) will only take
place if such difference is greater than the ratio
\(\frac{\sum_{p'\in P}F^{p'\text{,QF}}}{D}\) .

\begin{observation}

The marginal cost of reallocating funds by contributing to a project is $\frac{\sum_{p'\in P}F^{p'\text{,QF}}}{D}$, so this cost increases as more funds are contributed. 

\end{observation}

Indeed, we can see this latter ratio (the RHS in Equation
\ref{eq:sumwaverage}) as the marginal cost of contributing to reallocate
limited funds. This design has the property that the cost of
reallocating funds increases as more contributions are made.

It is also useful to rewrite Equation \ref{eq:sumwaverage} as
\begin{equation}
\sum_iV_i^{'p}(.)=\frac{\sum_{p'\in P}F^{p'\text{,QF}}}{D}+\sum_{p'\in P}\frac{F^{p'\text{,QF}}}{(\sum_{{p'}\in P}F^{p'\text{,QF}})}\sum_iV_i^{'p'}(.)
\label{eq:sumofvs}\end{equation}

Notice that the LHS of Equation \ref{eq:sumofvs} is the sum of the
marginal utility across individuals of increasing the size of the public
good \(p\). Also notice that the RHS is constant for every project
\(p\). Notice that Equation \ref{eq:sumofvs} implies \[
\sum_iV_i^{'p}(.)=\sum_iV_i^{'p'}(.) \text{ }\forall p,p'
\] Which is the condition for the socially optimal allocation for
limited funds.

\begin{observation}

The mechanism tends to equalize the sum of marginal utilities across public goods.

\end{observation}

Once the sum of marginal benefits are equalized enough across projects,
there are no further incentives to invest (n.b. under limited funds
contributors pay to reallocate funds but total funds are not further
increased). Indeed, by Equation \ref{eq:sumwaverage} contributions will
only take place as reallocating funds cover the cost
\(\frac{\sum_{p'\in P}F^{p'\text{,QF}}}{D}\).

This result implies another powerful property of QF. In addition to the
efficiency result in BHW, where the mechanism requires enough matching
funds to efficiently allocate funds for all projects, this shows that
the mechanism can also stimulate a social efficient allocation on the
margin with available funding.

\hypertarget{notes}{%
\subsection*{Concluding notes}\label{notes}}
\addcontentsline{toc}{subsection}{Concluding notes}

One of the most notorious characteristics of QF is that this mechanism
is able to solve the planner's problem in a decentralized way. Even a
benevolent, efficient social planner, will face the problem of knowing
the preferences of the community. By allowing the community to signal
their preferences, there seems to be an obvious advantage for a
decentralized arrangement.

Buterin, Hitzig, and Weyl (2019) (BHW) show that, when there are no
limits on matching funds, the decentralized allocation achieves the
socially efficient allocation. In a more realistic, limited funds
scenario, a question of interest is whether this mechanism achieves an
socially optimal allocation of limited funds. In the variation of the
mechanism we have proposed here, we have shown that this is the case.

It is worth mentioning that here we have assumed full information, but a
decentralized arrangement as the one that we have examined here, might
face some coordination problems related to available information. An
individual contributor invests on the basis of the contributions that
have been made by others, and their expectation on the final
distribution of the matching pool. Therefore such a contributor needs to
learn about the preferences of the community as they evaluate when and
how much to contribute.

With limited matching funds (in a realistic, limited-information case),
some contributions could have the unpleasant return of increasing the
public good of interest in less than the money invested. This seems to
imply that having information on the preferences of the community will
be even more valuable in this case. Also, with limited funds, the
mechanism has an auction flavor, where the cost of reallocating funds
increases as the round progresses.

If contributors learn from the contributions of others as the round
unfolds, increasing the amount of funds in the donor pool might provide
more time for learning without incurring costs. On the good side, all
allocated funds to the mechanism will be part of a new round of
financing, as part of the new round matching pool, so inefficient
investments could be re-distributed.

\hypertarget{references}{%
\section*{References}\label{references}}
\addcontentsline{toc}{section}{References}

\hypertarget{refs}{}
\leavevmode\hypertarget{ref-buterin_flexible_2019}{}%
Buterin, Vitalik, Zoë Hitzig, and E. Glen Weyl. 2019. ``A Flexible
Design for Funding Public Goods.'' \emph{Management Science} 65 (11):
5171--87.

\leavevmode\hypertarget{ref-pasquini_quadratic_2020}{}%
Pasquini, Ricardo A. 2020. ``Quadratic Funding and Matching Funds
Requirements,'' September, 49.
\url{https://doi.org/https://dx.doi.org/10.2139/ssrn.3702318}.

\bibliographystyle{chicago}

\bibliography{Bibliography-MM-MC}

\end{document}